\documentclass[12pt]{article}
\usepackage{amsfonts}
\usepackage{amsmath}
\usepackage{epsfig}

\begin{document}

\begin{flushright}
02/2009\\
\end{flushright}
\vspace{20mm}
\begin{center}
\large {\bf Some Remarks on the Tetron Spin Problem}\\
\mbox{ }\\
\normalsize
\vskip3cm
{\bf Bodo Lampe} \\              
e-mail: Lampe.Bodo@web.de \\   
\vspace{1.5cm}
{\bf Abstract}\\
\end{center}
The possible spatial transformation properties 
of tetrons are discussed.

\newpage

\section{Introduction} 

In recent papers \cite{lampe2,lampe3} the inner symmetries 
of the tetron model have been discussed on the basis of the 
known representations $A_1$, $A_2$, $E$, $T_1$ and $T_2$ 
of the permutation group $S_4$. 

Correspondingly, a $S_4$-'flavor' permutation index taking the 
values a,b,c and d has been introduced and 
suggested to be part of a larger continuous 
flavor symmetry SU(4). 
More concretely, it has been shown that the 24 $S_4$ 
representation states can be considered as part of 
the 256 SU(4)-states of the tensor product 
\begin{equation} 
{\bf 4\otimes 4\otimes 4\otimes 4}=3{\bf \times 45(T_1)}+3{\bf \times 15(T_2)}
                 +2{\bf \times 20(E)}+{\bf 35(A_1)}+{\bf 1(A_2)}
\label{eq1sdd}
\end{equation}
where one finds in brackets, which kind of $S_4$ symmetry 
functions are 
contained in the corresponding SU(4) representations, giving 
rise to the observed 3-generation spectrum of 
quarks ($T_1$ and $T_2$) and leptons ($A_1$, $A_2$ and $E$).

One then needs an exclusion principle which demands, that 
among the 256 SU(4)-states only the 24 $S_4$ representation 
states are physical. We shall see later how this 
exclusion principle may be related to the spatial 
transformation properties of tetrons which will 
be discussed now.

\section{The Tetron Spin Problem} 

Eq. (\ref{eq1sdd}) presupposes a constituent picture 
where quarks and leptons are built from 4 tetron 'flavors' 
a, b, c and d. In this secenario, one immediately faces the problem, 
how the spin-$\frac{1}{2}$ fermion behavior of quarks and 
leptons arises on a microscopic level. 

In this note I will present the framework, in which this 
problem should be solved. I will consider spatial transformations only. 
The extension to Minkowski space will be worked 
out in a separate publication \cite{lampef}.

Let me start with a few well-known facts about half-integer 
spin: in a physical experiment one cannot distinguish 
between states which differ by a complex phase. Therefore, in 
addition to ordinary representations one may include 
projective, half-integer spin representations of 
the rotation group SO(3), and also of its $T_d=S_4$ 
subgroup\footnote{$T_d$ is the rotation symmetry group 
of a regular tetrahedron. It is a subgroup of $O(3)$ 
and isomorphic to $S_4$. 
$T_d$ is also isomorphic to the octahedral 
group $O$, i.e. the group of proper rotations of a 
cube which is a subgroup of SO(3).}. These 
are true representations of the corresponding covering 
groups SU(2) and $\tilde{S_4}$, respectively.

To solve the tetron spin problem I suggest to give up 
the requirement of continuous rotation symmetry and assume that 
tetrons live and interact in microscopical environments, in which 
only permutation symmetry survives. The latter is 
much less restrictive than rotational SO(3), because 
the idea of rotation assumes concepts of angle and length, 
which may be obstacled by quantum fluctuations when approaching the 
Planck scale. In contrast, the idea of permutation merely 
presupposes the more fundamental principle of identity. This is why 
permutation groups may enter theoretical physics at finer 
levels of resolution and higher energies than the Lorentz group. 
Tetrons may be more basic than spinors. 

I call this assumption the 'spatial permutation hypothesis'. 
It amounts to introducing a second permutation index 
taking values 1,2,3 and 4 (in addition to the 
flavor index a,b,c and d) and being responsible for the 
spatial ('spin') transformation behavior of tetrons 
and its compound states. 

It is true that the phenomenological observation of 24 quarks 
and leptons and their interactions imply a permutation 
principle only on the level of inner symmetries (as in eq. 
(\ref{eq1sdd})). However, the assumption of 4 different tetron 
'spins' within a fermion bound state comes closest to the original 
intuition of a spatial tetrahedral structure as 
discussed in ref. \cite{lampe2} where a generic ansatz 
for the composite wave function $a_i b_j c_k d_l$ 
with $i,j,k,l \in \{1,2,3,4\}$ has been proposed. 

As a consequence of the spatial permutation hypothesis 
a new type of particle statistics 
will arise (called {\it tetron statistics}) 
which differs from Fermi and Bose statistics and is 
related to the 
exclusion principle in the flavor sector as formulated above.  


\section{The Details} 

According to the spatial permutation hypothesis, 
the spin part of a 4-particle fermionic compound state 
should transform according to a (projective) representation of $S_4$. 
Besides the ordinary representations $A_1$, $A_2$, $E$, $T_1$ and $T_2$
there are 3 irreducible projective representations (representations 
of the covering group $\tilde{S_4}$), namely 
$G_1$, $G_2$ and $H$ of dimensions 2, 2 and 4, respectively \cite{johnson}. 
The sum 4+4+16 of the dimensions squared accounts for the 24 additional 
elements due to the $Z_2$ covering of $S_4$.
Among them, $G_1$ uniquely corresponds to spin-$\frac{1}{2}$, i.e. is obtained 
as the restriction of the fundamental SU(2) representation to $\tilde{S_4}$. 
Similarly, $H$ can be obtained from the spin-$\frac{3}{2}$ representation 
of SU(2), whereas $G_2$ is obtained from $G_1$ by reversing the sign 
for odd permutations and contains contributions from spins larger 
than $\frac{3}{2}$. 

{\footnotesize 

For the understanding of the following arguments a 
short digression on quaternions and its usefulness for 
describing nonrelativistic 
spin-$\frac{1}{2}$ fermions will be helpful: 

Quaternions  
are a non-commutative extension of the complex numbers 
and play a special role in mathematics, because 
they form one of only three finite-dimensional 
division algebra containing the real numbers as a subalgebra. 
(The other two are the complex numbers and the octonions.)
As a vector space they are generated by 4 basis elementes 1, I, J and K 
which fulfill $I^2=J^2=K^2=IJK=-1$, where K can be obtained as a 
product $K=IJ$ from I and J. 
Quaternions are non-commutative in the sense IJ=-JI. 
Any quaternion q has an expansion of the form 
\begin{eqnarray}  
q&=&c_1+J c_2 \nonumber \\
 &=&r_1+Ir_2+Jr_3+Kr_4 
\label{eq302ui}
\end{eqnarray}
with real $r_i$ and complex $c_1=r_1+Ir_2$ and $c_2=r_3-Ir_4$.

There is a one-to-one corresponence between unit quaternions 
$q_0$ and SU(2) transformation  
matrices, because the latter are necessarily of the form 
$(\alpha, \beta; -\beta^*, \alpha^*)$ 
with complex $\alpha$ and $\beta$ fulfilling 
$|\alpha|^2 + |\beta|^2 =1$, and can be written as 
$q_0=\alpha + J \beta$. 
Therefore, the action of SU(2) matrices on spinor fields $(c_1,c_2)$ 
($c_1$ with spin up and $c_2$ with spin down) can 
in quaternion notation be rewritten as: 
\begin{equation} 
c_1+J c_2 \rightarrow (\alpha + J \beta)(c_1+J c_2)
\label{eq20511}
\end{equation}
For example the unit quaternions I and J corresponding to 
rotations by $\pi$ about the x and y-axis amount to  
$c_1 \rightarrow Ic_1, c_2\rightarrow -Ic_2$ 
and $c_1 \rightarrow -c_2, c_2\rightarrow c_1$, 
respectively. 
For a general SU(2) transformation one has  
$c_1 \rightarrow \alpha c_1-\beta^* c_2$ and 
$c_2 \rightarrow \alpha^* c_2+\beta c_1$, 
from which e.g. the antisymmetric 
tensor product combination $c_1 c_2' -c_2 c_1'$ 
can be shown to be rotationally invariant (spin 0). 
} 

To describe spin-$\frac{1}{2}$ bound states one should use 
the symmetry function of the representation $G_1$. This function will 
also be called $G_1$ in the following and can be given 
as linear combination of the $G_1$ representation 
matrices (=unit quaternions): 
\begin{eqnarray}  
G_1&=&  g(1,2,3,4)+  U g(2,3,1,4)+ U^2 g(3,1,2,4)  \nonumber \\
 &+&I g(2,1,4,3)+    S g(3,2,4,1)+ R^2 g(1,3,4,2) \nonumber \\
&+& J g(3,4,1,2)+    R g(1,4,2,3)+ T^2 g(2,4,3,1) \nonumber \\
&+& K g(4,3,2,1)+    T g(4,1,3,2)+ S^2 g(4,2,1,3) \nonumber \\
&+& \frac{I+K}{\sqrt{2}} g(3,2,1,4)+\frac{I-J}{\sqrt{2}} g(1,3,2,4)        
                                     +\frac{J+K}{\sqrt{2}} g(2,1,3,4)  \nonumber  \\
&+&  \frac{1-J}{\sqrt{2}} g(2,3,4,1)+\frac{1-K}{\sqrt{2}} g(3,1,4,2)
                                     +\frac{J-K}{\sqrt{2}} g(1,2,4,3) \nonumber   \\
&+&  \frac{I-K}{\sqrt{2}} g(1,4,3,2)+\frac{1+K}{\sqrt{2}} g(2,4,1,3)
                                     +\frac{1+I}{\sqrt{2}} g(3,4,2,1) \nonumber   \\
&+&  \frac{1+J}{\sqrt{2}} g(4,1,2,3)+\frac{I+J}{\sqrt{2}} g(4,2,3,1)
                                     +\frac{1-I}{\sqrt{2}} g(4,3,1,2)  
\label{eq3029}
\end{eqnarray} 
where $R=\frac{1}{2}( 1- I- J- K), 
S=\frac{1}{2}( 1- I+ J+ K), T=\frac{1}{2}( 1+ I- J+ K)$ 
and $U=\frac{1}{2}( 1+ I+ J- K)$. 
One can see explicitly from this equation, 
which $S_4$ permutation $\overline{ijkl}$ is represented 
in $G_1$ by which quaternion, because the corresponding 
quaternion appears as a 
coefficient of g(i,j,k,l). For example, the permutation $\overline{2341}$ 
is represented by $\pm (1-J)/\sqrt{2}$, and so on. 
In other words, the quaternion coefficients $1, I, J, K, 
(I+K)/\sqrt{2}, ..., R, S, T, ...$ in this equations 
represent the elements of $\tilde{S_4}$\footnote{ 
While $\tilde{S_4}$ itself can be shown to make up the inner shell 
of $D_4$-lattices \cite{dixon}, 
the first half of coefficients in eq. (\ref{eq3029}) represent 
even permutations corresponding to $\tilde{A_4}$ 
which is sometimes called the 'binary tetrahedral group', 
and generates the $F_4$ lattice also called the ring of 
Hurwitz integers (=quaternions with half integer coefficients). 
The Hurwitz quaternions form a maximal order (in the sense of ring theory) 
in the division algebra of quaternions with rational components. This 
accounts for its importance. For example restricting to integer 
lattice points, which seems a more obvious candidate for 
the idea of an integral quaternion, one does not get a maximal order 
and is therefore less suited for developing a theory of left 
ideals as in algebraic number theory. 
What Hurwitz realized, was that his definition of integral quaternions 
is the better one to operate with. 
}. 

Due to the 2-fold covering of $S_4$ each of the 
real functions $g(i,j,k,l)$ 
in eq. (\ref{eq3029}) with its 24 terms is in fact a difference 
$p(i,j,k,l)-m(i,j,k,l)$ so as to obtain the 48 terms needed for 
a symmetry function of $\tilde{S_4}$. 

Eq. (\ref{eq3029}) should be considered as the spin factor of the 
4-tetron bound state, whereas the $A_1$, $A_2$, $E$, $T_1$ 
and $T_2$-functions of the ordinary $S_4$ representations 
account for the flavor 
factor. In fact, working out the quaternion multiplications in 
eq. (\ref{eq3029}) and using $K=IJ$ one obtains a representation 
of the form $G_1= c_1+J c_2$ 
with $c_1$ and $c_2$ decribing the 2 spin directions 
of the compound fermions, cf eq. (\ref{eq20511}). 
Mathematically, the appearance of 2 complex functions $c_1$ and $c_2$ 
in eq. (\ref{eq3029}) is merely expression of the fact that for the 2-dimensional 
representation $G_1$ 4 real(=2 complex) 
symmetry functions can be constructed, which in eq. (\ref{eq3029}) are 
combined in one quaternion function. 

Eq. (\ref{eq3029}) therefore describes a decent fermion state 
which transforms in the standard way, cf. eq. (\ref{eq20511}).
On the other hand, eq. (\ref{eq3029}) also 
inherits the spatial permutation hypothesis (i.e. giving up full 
SU(2) rotational invariance on the tetron level) in 
that the function $G_1$ naturally reacts like a (projective) $S_4$ 
representation under permutations of $i,j,k,l \in \{1,2,3,4\}$. 

Since it is not possible to build a spin-$\frac{1}{2}$ $G_1$ 
state as a 4-tensor product similar to eq. (\ref{eq1sdd}), 
the picture followed here is a sort 
of molecular approach where one starts with a fixed 
spatial tetrahedral configuration with 4 distinct 
permutation 'spin' indices $i,j,k,l \in \{1,2,3,4\}$ 
which according to the spatial permutation hypothesis 
must transform according to $G_1$. 
Its reaction under permutations ($T_d$ transformations) of i,j,k,l 
is dictated by the spatial permutation hypothesis, 
whereas the behavior under full rotational SU(2) is obtained from the 
requirement that the compound state must be a fermion. 

If one wants to go beyond this understanding one should look 
for possible transformation properties of the spatial 
permutation indices i,j,k,l in a tensor product, which mimics 
the behavior of $G_1$. Since this cannot be obtained 
within the usual framework of universal $Z_2$ coverings, 
one has to consider e.g. octonion $Z_4$ extensions of the rotation group. 
In such a framework 
$g(i,j,k,l)$ (or alternatively p and m) are the functions 
which should 
be interpreted as tensor products of tetrons of the generic form 
\begin{equation} 
g(i,j,k,l) = a_i\otimes b_j'\otimes c_k''\otimes d_l'''
\label{eq302rr}
\end{equation}
where a,b,c,d are the tetron 'flavor' and i,j,k,l their 'spin' indices, 
so that the complete spin and flavor wave function of quarks and leptons
can be written as
\begin{eqnarray}  
a_1\otimes b_2'\otimes c_3''\otimes d_4'''  &+& b_1\otimes c_2'\otimes a_3''\otimes d_4''' + ... \nonumber \\
I a_2\otimes b_1'\otimes c_4''\otimes d_3''' &+& ... \nonumber \\
...& &
\label{eq302}
\end{eqnarray}
Here in the rows the tetron flavor indices a,b,c,d are permutated in order 
to obtain the appropriate flavor combination ($A_1$ of $S_4$ as an example, 
for the $A_2$, $T_1$ etc flavor representations $G_2$ and $H$ will 
come into play), 
whereas in the columns the tetron spin indices i,j,k,l are permutated 
in order to obtain the $G_1$ spin combination.\footnote{
Note that in general, the permutation of the tensor product 
indices - denoted by primes in eq. (\ref{eq302rr}) - must not be 
messed up with the permutation of spin states. Only 
in the case at hand, where 4 different spin states in 
4 different tensor factors are considered, there is no difference. 
}

Very important: eq. (\ref{eq3029}) reflects the statistical behavior of a  
4-tetron conglomerate unter permutations of its components. 
This behavior has a 
certain similarity to that of fermions but is certainly not 
identical. While conglomerates of fermions usually transform 
with the totally antisymmetric representation (like $A_2$), 
tetrons go with $G_1$, which gives a factor of I under 
the exchange ($1\leftrightarrow 2,3\leftrightarrow 4$)
or $\frac{1}{\sqrt{2}} (J+K)$ under ($1\leftrightarrow 2$), 
whereas a 2-fermion conglomerate in a $A_2=c_1 c_2' -c_2 c_1'$ 
configuration responds with -1 (i.e. antisymmetric) to the exchange 
of ($1\leftrightarrow 2$). See table 1, where the behavior of 
tetrons and fermions is compared. The fact that tetrons behave 
more complicated under transpositions $(i \leftrightarrow j)$, 
has to do with the fact that transpositions in $S_4$ correspond 
to relatively complicated space transformations in $T_d$. 

We therefore conclude that tetrons follow their own statistics 
which is neither bosonic nor fermionic, and assert, that 
a sort of 'tetron spin statistics theorem' holds, which 
allows only bound states in which all tetron flavors are different. 
This then explains the selection rule/exclusion 
principle proposed in ref. \cite{lampe3} and 
mentioned after eq. (\ref{eq1sdd}).

To prove the assertion I can offer the 
following argument: remember that the Pauli principle 
for fermions demands antisymmetry ($A_2$ behavior) 
of a compound wave 
function under the exchange of all (spin and flavor) indices. 
In the case of tetrons one analogously needs a $G_1$ 
behavior of the compound wave function under the 
simultaneous permutations of all permutation indices 
(i,j,k,l and a,b,c,d), e.g. 
$a_i b_j' c_k'' d_l''' \rightarrow b_j a_i' c_k'' d_l'''$ 
for $(1\leftrightarrow 2) \in S_4$.
Such a behavior could not be obtained, if two or more 
flavor indices were identical. 

\begin{table}
\label{tab4}
\begin{center}
\begin{tabular}{|c|c|}
\hline
FERMIONS                             &  TETRONS   \\
\hline
\hline
\multicolumn{2}{|c|}{compound states:}                         \\
\hline
boson from 2 fermions:                &  fermion from 4 tetrons: \\
complex tensor product                &  quasi-complex, quaternion tensor product \\
$A_2=c_1 c_2' -c_2 c_1'$              &  $G_1=g(1,2,3,4)+I g(2,1,4,3)+J...$ \\   
                                      &  $=a_1 b_2' c_3'' d_4''' + I a_2 b_1' c_4'' d_3''' +...$ \\
bosonic behavior under rotations      &  fermionic behavior under rotations \\
                                      &  $G_1\rightarrow (\alpha + J \beta)G_1$ \\
\hline
\hline
\multicolumn{2}{|c|}{permutation behavior/statistics:}                         \\
\hline
-1 under ($1\leftrightarrow 2$)      &   a factor I under ($1\leftrightarrow 2,3\leftrightarrow 4$) \\  
                                     &   a factor $\frac{1}{\sqrt{2}} (J+K)$ under ($1\leftrightarrow 2$) etc \\
\hline
\end{tabular}
\bigskip
\caption{Comparison between the known fermion behavior and the anticipated 
tetron behavior.}
\end{center}
\end{table}

\section{Conclusions}

It is certainly true that 
the phenomenological observation of 24 quarks 
and leptons and their interactions suggest a permutation 
principle only on the level of {\it inner} symmetries. However 
due to the problems which arise in connection with spin and 
statistics one is naturally lead to consider the possibility 
that inner and outer permutation behavior may be intertwined 
and that this can be used to understand the spin-$\frac{1}{2}$ 
nature of quarks and leptons. 

If the tetron approach has 
some meaning it is possible that 
besides $G_1$ also the two other half-integer spin 
representations of $\tilde{S_4}$ ($H$ and $G_2$) play 
a role in nature, or in other words, that particles with spin 
$\frac{3}{2}$ and $\frac{5}{2}$ may appear at higher energy levels.

\end{document}